\documentstyle[prb,aps,multicol,epsfig]{revtex}    
\begin{document}    


\twocolumn[ 
\hsize\textwidth\columnwidth\hsize\csname@twocolumnfalse\endcsname 

\title{Dynamic scaling theory of the Kosterlitz-Thouless-Berezinskii transition: 
ubiquitous finite size effects} 

\author{Louis Colonna-Romano,$^*$ Stephen W.~Pierson,$^\dag$}   

\address{Department of Physics, Worcester Polytechnic Institute,
Worcester, MA 01609-2280}    

\author{Mark Friesen$^\ddag$}   

\address{University of Wisconsin, Madison, WI 53706}  

\date{Received \today}    
\maketitle    
\begin{abstract}    
A numerical study of the neutral two-dimensional (2D) lattice Coulomb gas is performed
to examine dynamic scaling, finite size effects, and the dynamic 
critical exponent $z$ of the Kosterlitz-Thouless-Berezinskii transition. 
By studying large system sizes ($L=100$), we show $z=2.0 \pm0.2$ using 
Fisher-Fisher-Huse (FFH) scaling. We also present evidence that the vortex 
correlation length is finite below the transition temperature, in contrast to 
conventional wisdom. Finally, we conclude that previous findings for a variety 
of experimental systems that $z\simeq 5.6$ using FFH scaling indicates that 
finite size effects are ubiquitous in 2D superconductors and Josephson Junction
Arrays. 

\end{abstract}    
] 
\narrowtext    

\section{Introduction}    
\label{sec:intro} 

The universality class of the two-dimensional (2D) XY model 
has been fascinating from the start with its phase transition 
being driven by topological excitations, the 
unusual, exponential temperature dependence of the correlation length,
and its apparent presence in systems ranging from 2D superfluids 
and superconductors to liquid crystals and magnetic 
systems.\cite{kt73,k74,b71} Despite its uniqueness, the detection 
of the Kosterlitz-Thouless-Berezinskii (KTB) 
transition,\cite{kt73,k74,b71} characterized by an unbinding of vortex
pairs, has been challenging, but several techniques have been developed to 
probe its critical behavior. In two-dimensional superconductors and Josephson 
Junction arrays (JJA's), an examination of the dc transport properties in zero 
magnetic-field has been the primary method, but the 
magneto-resistance\cite{garland87} and the ac kinetic 
inductance\cite{fiory83} have also been used. More 
recently, magnetic flux noise measurements\cite{shaw96} have 
been used to study the KTB transition. And, as we discuss shortly,
numerical methods and scaling approaches have also been 
used. The advantage of the scaling methods is that they
can reveal information on the dynamic critical exponent $z$,
one of the key exponents in the KTB critical behavior, which 
relates the relaxation time $\tau$ to the vortex correlation 
length $\xi$ via $\tau\propto\xi^z$.

While there are many methods for studying the KTB 
transition,\cite{minnhagen87} 
the most common is the ``conventional" method which involves 
an analysis of the dc voltage-current ($V$-$I$) isotherms and 
the temperature dependent resistance $R(T)$. In this 
approach, the transition temperature $T_c$ is determined
via the $V$-$I$ exponent $\alpha(T,I)$ ($V\propto I^\alpha$), 
which is predicted to have a value of 3 at $T_c$ (assuming $z=2$,) 
and whose value should decrease rapidly for weak currents near 
$T_c$. The Ohmic resistance is then fitted by the  
Kosterlitz\cite{k74} or Minnhagen form\cite{minnhagen87} for the 
resistance, which are respectively, $R\propto \exp [-bz/\sqrt{T-T_c}]$ 
and $R\propto \exp [-bz\sqrt{(T_{c0}-T_c)/(T-T_c)}]$, where $b$ is a 
system-specific parameter and $T_{c0}$ is the mean-field transition 
temperature. 

There are many weaknesses to the conventional approach.\cite{pierson99} 
For example, it is valid only in the weak-current
limit where few experiments can probe effectively and where finite-size 
effects dominate. Moreover, the exponent $\alpha$ has a current dependence
and so determining its value uniquely for a given temperature is not possible.
In fact, researchers can use this flexibility to their advantage to show
a rapid variation of $\alpha$ with temperature near $T_c$. This is done 
by determining $\alpha$ in different current ranges for each temperature
so as to achieve the desired behavior. (The most accurate method would be 
to determine $\alpha$ for the $V$-$I$ 
isotherms for the same current so as to represent a single length 
scale.\cite{pierson99}) Another problem with the traditional approach 
is the number of fitting parameters in the resistance formulas, which 
allow excessive freedom for interpreting results. For these reasons, 
it is important to examine other approaches to verifying and studying 
KTB behavior.

The Fisher-Fisher-Huse (FFH) dynamic scaling,\cite{ffh91} which says that
$V\xi^z/I$ is a function of $I\xi/T$, has its advantages over the conventional
approach. The most prominent is that it is valid over an extended current range
and therefore incorporates the current dependences of $\alpha(T,I)$. Moreover,
FFH scaling can be used to determine $z$. Recently, it has been 
used to study experimental, dc $V$-$I$ data but yields a large value of the 
dynamic critical exponent $z\simeq 5.6$ for a wide variety of samples and 
systems.\cite{pierson99,ammirata99,pierson00b} It has recently been reiterated
that the FFH scaling is only valid in the absence of finite size effects and 
suggested that finite size could result in large values of $z$.\cite{minnhagen00}

Finite size effects, which include a finite size system and the inherent
magnetic penetration depth $\lambda_\perp$\cite{pearl64} in superconductors, have 
long been known to complicate the verification of KTB behavior because
they make free vortices possible for any temperature range. Kadin {\it et 
al.}\cite{kadin83} and Lee and Garland\cite{garland87} were among the 
first to study this experimentally while Repaci {\it et al.}\cite{repaci96} 
and Herbert {\it et al.}\cite{herbert98} have recently extended this work.
Theoretically, Simkin and Kosterlitz\cite{simkin97} have investigated 
numerically and analytically the finite size effects in these systems, while
Artemov\cite{artemov98} has examined the effect of ``Pearl'' vortices on
the magnetization in superconducting thin films. And more recently,
Pierson and Valls\cite{pierson00a} have completed a renormalization group
study of the two-dimensional Yukawa Gas, which incorporates the inherent
finite size $\lambda_\perp$ into the two-dimensional Coulomb Gas (2DGG).

Finite size (FS) scaling, which incorporates the effect of a finite size into the FFH
scaling, is an established method in the study of numerical data in KTB behavior. 
Lee and Teitel\cite{leeteitel94} were among the first to use this approach to show 
that the dynamic critical exponent $z$ has a value $2$ for the two-dimensional Lattice 
Coulomb Gas (2DLCG). Medvedyeva {\it et al.}\cite{minnhagen00} have also used finite 
size scaling to reinforce the result $z=2$ for the 2D resistively-shunted-junction 
model. (Numerical studies
in fact have provided many insights into KTB behavior. A review of those results
is beyond the scope of this paper but we refer the reader to many excellent
papers in the literature.\cite{jose97,leeteitel92,holmlund96,weber96,jonsson97,kim99})

Like the conventional approach and FFH scaling, FS scaling has its strengths and 
weaknesses. For example, because FS scaling requires the ability to carefully tune 
the system size, it is not conducive to studying experimental data where the finite
length scales are more illusive.  It is therefore important to keep in mind that all 
approaches provide a great deal of information and often complement one another.

The correlation length is an important quantity in FFH scaling but questions about its 
form below $T_c$ remain. In the original paper of Kosterlitz,\cite{k74} the correlation 
length was taken to be infinite below the transition temperature owing to the 
divergence of the susceptibility there. In subsequent papers by 
Friesen,\cite{friesen95a} Pierson,\cite{pierson95a} and Simkin and 
Kosterlitz,\cite{simkin97} the vortex correlation length is taken to be finite 
for this temperature regime (but still diverging at $T_c$ from below) 
and interpreted roughly as the size of the largest vortex pairs. Other 
authors\cite{minnhagen00} maintain that the correlation length is 
infinite for $T<T_c$. 

In this paper, we do a numerical study of the 2DLCG using Monte Carlo 
simulations to understand the difference in the value of $z$ obtained
from finite size scaling and FFH scaling. We vary the system size to 
study the effect of finite sizes and verify that $z=2$ using FFH
scaling, thereby making FFH scaling consistent with FS scaling. 
We also analyze the transport characteristics and find evidence that
the vortex correlation length is finite below the phase transition, in 
contrast to the claims of others.\cite{minnhagen00} The numerical approach 
that we use is that of Lee and Teitel\cite{leeteitel94,leeteitel92} 
extended to larger system sizes ($L=200$).

The paper is organized as follows: Section II describes the model that 
we use along with the numerical and scaling methods. In Section III,
we present our results and analysis of the transport characteristics.
Implications and a summary of our work is presented in Section IV.

\section{Model and Methods} 
\label{sec:model} 

\subsection{2D Lattice Coulomb Gas with an applied field}

To study the KTB critical behavior and its dynamic critical exponent,
we use the 2DLCG model, whose relationship to the KTB transition and mapping 
to the 2DXY model are well known.\cite{leeteitel94,xymap} This model 
is described in Refs.~\onlinecite{leeteitel94} and \onlinecite{leeteitel92} 
and so will be only briefly described here.

The Hamiltonian for the 2DLCG is given by 
\begin{equation}
\label{cgham} 
H = \frac{1}{2} \sum_{i,j} q_i V({\bf r}_{ij}) q_j,
\end{equation} 
where the sum is over all pairs of sites in the lattice, $q_i$ is 
the (integer) charge at site $i$, ${\bf r}_{ij} = {\bf r}_i -
{\bf r}_j$, and ${\bf r}_i$ is the location of the $i$th charge.  
$V$, which solves the discrete form of Poisson's equation in two dimensions,
is the Coulomb potential (and is therefore logarithmic at large distances). 
Using Fourier tranforms, $V$ in Eq.~(\ref{cgham}) may be replaced by 
\begin{equation}
V({\bf r}) = \frac{1}{N} \sum_k V_k (e^{i {\bf k}\cdot{\bf r}} - 1)
\end{equation} 
where, for the square lattice considered here,
\begin{equation} 
V_k = {\pi \over {2 - \cos ({\bf k}\cdot{\hat {\bf x}}) -
\cos ({\bf k}\cdot {\hat {\bf y}})}}. 
\end{equation}
The wave vectors are defined in the usual way: 
\begin{equation} 
{\bf k} = {{2 \pi m_1}\over L}
{\hat {\bf x}} + {{2 \pi m_2}\over L} {\hat {\bf y}}, 
\end{equation} 
where $m_1, m_2 = 0, 1, 2,\dots,L-1$.  In addition, since we wish
to apply our results to 2D superconductors in zero-magnetic field, we use
the constraint that the total charge in the system must be zero, which 
deals with the infrared divergence as ${\bf k} \rightarrow 0$.  

In the presence of an electric field, ${\bf E}$, an electric potential energy term 
of the form $\sum_i q_i {\bf r}_i\cdot{\bf E}$ is added to the 
Hamiltonian.\cite{leeteitel94} 
The new term results in a net movement of charge along the direction of the field ${\bf E}$
that we measure in terms the charge current density, ${\bf J}$, using the approach of
Ref.~\onlinecite{leeteitel94}. 

The applied field $E$ and the resulting vortex current density $J$ in the 2DLCG model 
are analogous to the current $I$ and voltage $V$ in a superconducting thin film, 
but in an inverted way due to the mapping between the two systems. In the 
superconducting thin film, the current is injected
into the sample, which causes vortex movement in a direction perpendicular to the 
current via the magnus force. This results in resistance and therefore a voltage $V$. 
For the 2DLCG, it is the voltage that is applied and that results in the vortex ``charge'' 
current in a direction parallel to the electric field. Following the notation  of Lee and 
Teitel,\cite{leeteitel94} the following correspondences between the two systems apply: 
$E\longleftrightarrow I$ and $J\longleftrightarrow V$. In this paper, we will examine the 
$E$-$J$ curves of the 2DLCG, which correspond to $V$-$I$ characteristics in actual 
superconducting thin films. Unless otherwise noted, the notation that we use will be 
that of the 2DLCG. Along these lines, the $E$-$J$ curves will be plotted as $E$ versus
$J$ in our figures, so that they mirror the analogous curves in the 2D superconcuctors
that are plotted as $V$ versus $I$. We will therefore use the notation $V$-$I$ for the 
current-voltage curves instead of the more standard $I$-$V$.

\subsection{Monte Carlo Simulations}

The Metropolis \cite{metropolis} Monte Carlo technique is used to
calculate the $J$-$E$ data based on the Hamiltonian discussion in Section IIA.  
Briefly, the technique, as it applies to the 2DLCG model with a square 
($N = L\times L$) lattice and integer charges, is as follows: (See
Ref.~\onlinecite{leeteitel94} for more details.)

{\parindent = 8 pt \narrower
\noindent 1.  Create a trial system from the original system by
choosing a site, $i$, and one of its four neighbors, $j$, at random.
Add one unit of charge to site $i$ and subtract one unit from site
$j$ so that the net charge in the system is always zero.

\noindent 2.  Compute $\Delta H = H_{trial} - H_{original}$.  Since only
two sites have been altered, most terms in $H_{trial}$ and $H_{original}$
are the same, simplifying the calculation of $\Delta H$.  

\noindent 3.  If $\Delta H < 0$ or if $e^{-\Delta H/T}$ is greater than a 
random number uniformly distributed on the interval $[0,1)$, accept the 
change and replace the the original system with the trial system for
subsequent trials.  Otherwise, leave the system unchanged.  \par}

When evaluating the Hamiltonian for a trial system, the minimum image
convention and toroidal (periodic) boundary conditions are used.\cite{allentilde}  
In addition, a ``stability'' term of the form $\sum_i (q^4_i - q^2_i)$ 
is added to the Hamiltonian to minimize the probability of creating
multiply charged sites in the system.\cite{leeteitel92} 

For many combinations of $E$ and $T$, particularly when the charge
density is small, a low rate of acceptance of trials is expected. When 
this occurs, it is convenient to calculate $\Delta H$ without summing over 
the array. The method employed here, used and described by Lee 
and Teitel,\cite{leeteitel92} is to store information on each site in the 
system in the form of the potential, which is updated for each site after each 
accepted trial. This technique can improve run times on the order of the 
reciprocal of the acceptance ratio, often factors of $O(10^3)$ or more. 

A sequence of $N$ trials is called one Monte Carlo Step (MCS).  
All of our computer runs consist of an initial $10,000$ MCS
to equilibrate the system during which no data are saved, followed by at
least $500,000$ MCS, for which the average current is computed.  The only
exceptions were some large systems ($L=200$) at $E > 0.6$ where a smaller
number of MCS was used due to very long run times.  At high electric
field, the current is large and the shorter runs provide sufficient
accuracy.  At each $E$ and $T$, at least five runs were performed.  The
error bars on the points in Figure \ref{L24raw}, Figure \ref{finsiz}, and
Figure \ref{L100raw} are the standard deviations of the average current
from these runs.  In some cases at low $E$, many additional runs were
performed in an attempt to reduce the error bars.  

\subsection{Fisher-Fisher-Huse and Finite-Size Scaling Methods}

Given the difficulties of the conventional approach to analyzing $V$-$I$ characteristics 
discussed in Section \ref{sec:intro}, it is important to use other methods including 
the powerful scaling methods. In this paper, we will use Fisher-Fisher-Huse scaling and 
finite size scaling.

In the FFH theory,\cite{ffh91} for 2D superconductors, 
the $E$-$J$ (in the notation of the 2DLCG) curves should scale as
\begin{equation}
\label{ffheq}
J=E\xi^{-z} \chi_\pm(E\xi/T), 
\end{equation}
where $\chi_{+ (-)} (x)$ is the scaling function for temperatures 
above (below) $T_{KTB}$.  
The two important asymptotic behaviors of $\chi(x)$ are
$\lim_{x\rightarrow 0} \chi_+ (x)= \mbox{const.}$ (Ohmic limit), 
and $\lim_{x\rightarrow \infty}\chi_\pm (x)\propto x^z$ 
(critical isotherm). As established in previous 
papers,\cite{pierson99,ammirata99,pierson00b}
it is advantageous to rewrite  Eq.~(\ref{ffheq}) as  
\begin{equation}
\frac{E}{T}\Biggl(\frac{E}{J}\Biggr)^{1/z}=\varepsilon_\pm(E\xi/T)
\label{FFHsc}
\end{equation}
(where $\varepsilon_\pm (x)\equiv x/\chi_\pm^{1/z} (x)$) because only 
the $x$-scale is stretched in Eq.~(\ref{FFHsc}). In Eq.~(\ref{ffheq}), 
both the $x$-scale and $y$-scale are stretched making it harder to 
judge a collapse of the scaled data. 

The FFH scaling has three fitting parameters $z$, $T_c$, and $b$, which 
can be determined in the three scaling steps. The first, which 
yields estimates of $z$ and $T_c$, is to determine the critical 
isotherm. In the absence of finite size effects, the critical isotherm
is straight on a log-log plot since $J\propto E^{z+1}$ for all $E$ at 
$T=T_c$. The second step is to estimate $b$ in the exponent of the correlation 
length, which, for experimental data, one can do by examining the 
resistance since $R(T)\propto \xi^{-z}.$ For numerical data, one assumes 
$\xi\propto\exp[C/\vert T-T_c\vert^{1/2}]$,\cite{pierson00b,leeteitel94} 
which is generally expected to hold for $T>T_c$.
The final step is to achieve a scaling collapse for Eq.~(\ref{FFHsc}), adjusting
the three parameters within the error bars established in steps 1 and 2. One 
should also bear in mind that all these steps can be repeated to optimize 
the scaling collapse.

In the finite size scaling form, valid for $\xi>L$, the system size $L$ is 
the dominant length scale and therefore substitutes for $\xi(T)$ in 
Eq.~(\ref{ffheq}) to arrive at\cite{leeteitel94}
\begin{equation}
\label{fsseq}
JL^{z}/E = \chi_\pm(EL). 
\end{equation}
Strictly, this equation works only for $T=T_c$, but is also valid for $T\simeq T_c$
since leading order corrections are small: $O(L/\xi)$. 
In this powerful approach, there is only one fitting parameter, $z$, making its
results more definitive.

To achieve the best scaling collapse for the FFH and FS scaling, the principle 
method that we use is to sort the data according to the value of the scaling 
variable $x$ (whether it be $E\xi/T$ or $EL$) and then to minimize the difference 
of the logarithm of the values of two adjacent, scaled points using a least squares 
method.\cite{pierson99} As a second means of optimization, we tried approximating 
the $T>T_c$ curve to a function and then minimizing the difference between the 
scaled points and the curve. However, because of the large number of fitting 
parameters, definitive results could not be achieved with this approach.

For various reasons, not all of the data are used in the optimization.
At large values of the applied electric field ($E\simeq 0.8$), the isotherms 
become ohmic due to a saturation of the vortex density. (See discussion in 
Section \ref{sec:results}A.) Because this behavior is not due to critical 
behavior, the data should not scale (collapse) with the other data and so is 
not included in our optimization. However, we do plot those points in the scaling 
plots (Figures \ref{L24FFH} and \ref {L100FFH}), even though those points do not 
collapse. (In Refs.~\onlinecite{pierson99,ammirata99} and \onlinecite{pierson00b},
the analogous, large $I$ data was not plotted with the scaled data.) In addition,
because the dynamic critical exponent $z$ has a temperature dependence below 
$T_c$ due to the line of fixed points in the KTB critical behavior, the $T<T_c$
data are not used in the optimization of the FFH scaling collapse. This is 
because the temperature dependence of $z$ is not the same as the scaling 
variable $x=E\xi/T$ and so the FFH scaling is not expected to be as 
effective in this regime.\cite{pierson99,minnhagen00} Nonetheless, we do scale
the $T<T_c$ data assuming a symmetric correlation length. Finally, data are also 
excluded from the optimization if the error bars are greater than $50\%$.

\section{Results}
\label{sec:results}

The principal objectives of our work are to study the value of the dynamic 
critical exponent $z$ for the 2DCG in the context of Fisher-Fisher-Huse 
scaling and the influence of finite size effects, and to study the behavior 
of the correlation length $\xi_-(T)$ below the phase transition. 

\begin{figure}[t]
\centerline{
\epsfig{file=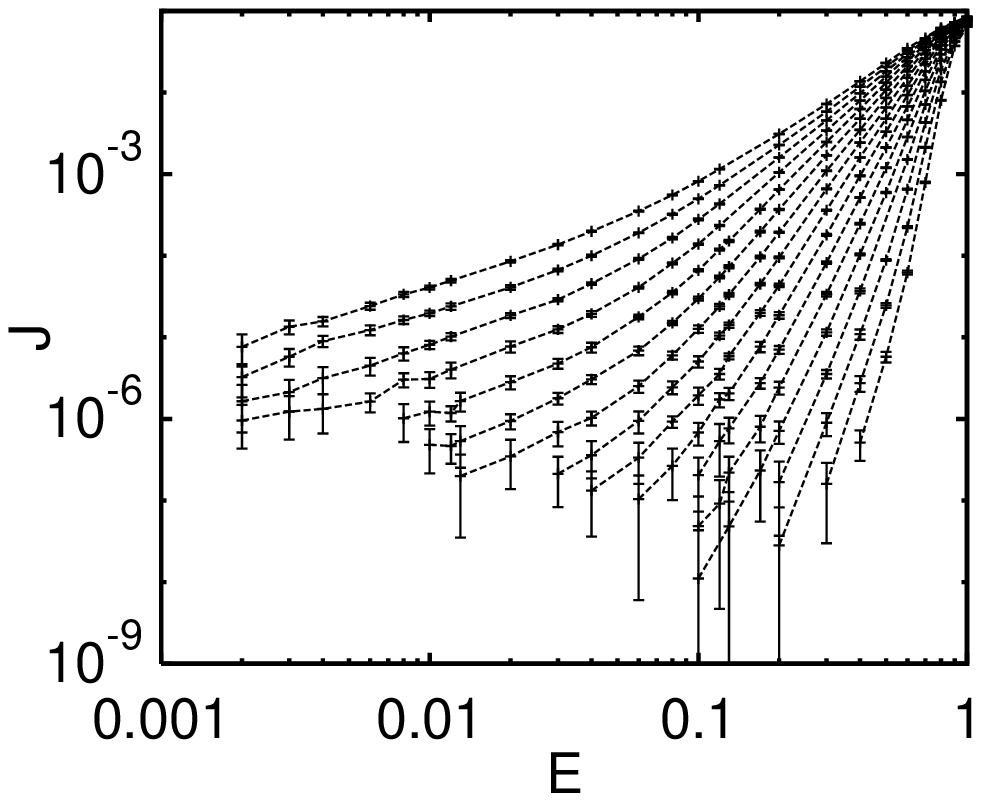,angle=0,width=3.4in}}
\caption{
Charge current density $J(E,T)$ versus $E$ for fixed system
size $L=24$ for temperatures ranging from 0.09 to 0.25 by increments of
0.01 (from bottom to top).}
\label{L24raw}
\end{figure}

\subsection{Dynamic Critical Exponent $z$}

The two principal system sizes used in our simulations were $L=24$ and 
$L=100$. The $J$-$E$ isotherms for the $L=24$ system are shown in Figure 
\ref{L24raw}. As one can see the isothermal $J$-$E$ curves mimic closely 
the $V$-$I$ of thin-film superconductors.\cite{repaci96} Reflecting the 
absence of vortex pairs, the higher temperature curves are primarily 
Ohmic (i.e., have slope 1 on a log-log plot). The lowest 
temperature curves are primarily non-Ohmic, indicating the presence of 
vortex pairs. The intermediate-temperature curves, on the other hand, are 
Ohmic for small $E$, become non-Ohmic at larger values of $E$, and 
abruptly turn Ohmic for the largest values of $E$. This of course 
reflects the fact that with each different value of $E$, one is probing a 
different length scale, which varies as the inverse of $E$.\cite{ahns80} (See
discussion in last paragraph of Section IIIB.)
Consequently, at small $E$, long length scales are being probed where free 
vortices exist and Ohmic behavior is expected for $T>T_c$ (or when 
$\lambda_\perp$ is finite). For large values of $E$, 
smaller length scales are probed where vortices are still bound in pairs 
and result in non-Ohmic behavior. The other salient feature of this graph 
is that the error bars become large at the small-$E$ end of the isotherms 
due to small numbers of vortex pairs.

\begin{figure}[t]
\centerline{
\epsfig{file=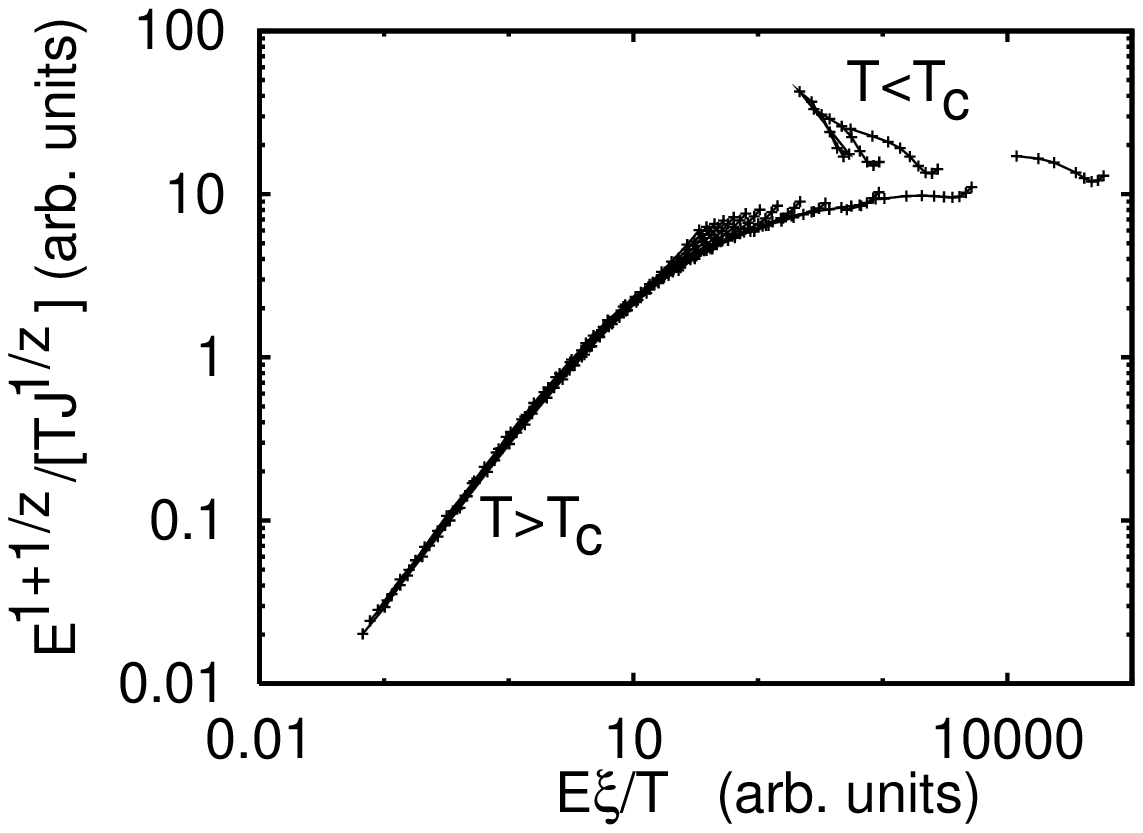,angle=0,width=3.4in}}
\caption{
The $L=24$ data of Figure \protect\ref{L24raw} scaled according
to FFH scaling (Eq.~\protect\ref{FFHsc}) using $z=6.01$, $T_c=0.127$, and $C=0.7475$.} 
\label{L24FFH}
\end{figure}

The temperature of the phase transition in the $L=24$ data is not obvious 
in Figure \ref{L24raw}. Using the FFH dynamic scaling criteria for the 
critical isotherm, which says that $J\propto E^{z+1}$ for all 
values of $E$ at $T=T_c$, results in $T_c\simeq 0.14$ and $z\simeq 6$. 
The subsequent FFH scaling for the $T>T_c$ data yields an excellent 
collapse for $z=6.01$, $T_c=0.127$, and $C=0.7475$, as seen in Figure 
\ref{L24FFH}. (As noted in Section \ref{sec:model}, the crossover
to Ohmic behavior at large $E$ data seen Figure \ref{L24raw} is not 
due to critical behavior and so should not be used to judge the collapse. It 
is this large-$E$ data that does not collapse with the other data in Figure 
\ref{L24FFH}.) 

\begin{figure}[t]
\centerline{
\epsfig{file=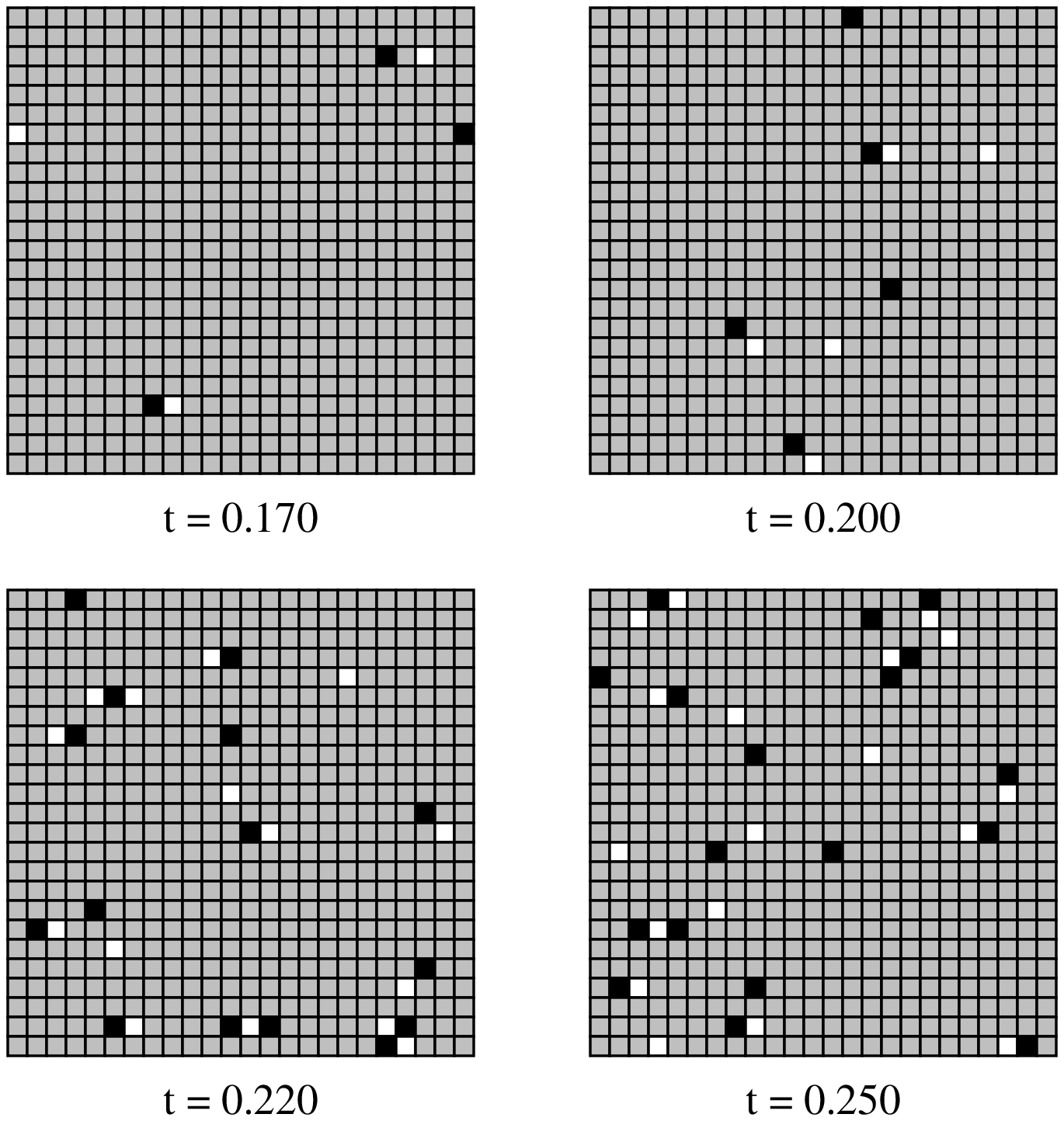,angle=0,width=3.4in}}
\caption{
Sample vortex configurations for $E=0.3$ at four temperatures. 
A white box is a vortex, a black box an anti-vortex, and a grey box an empty lattice site.} 
\label{dielectric}
\end{figure}

The values of the parameters in this collapse, optimized using our 
method described in Section \ref{sec:model}, are consistent with the 
large values of $z$ and low values of $T_c$ found 
previously.\cite{pierson99,ammirata99,pierson00b} However, an examination 
of the inverse dielectric constant\cite{leeteitel92} $\epsilon^{-1}$ 
and the vortex density\cite{minnhagen87} reveals no critical behavior at that temperature 
with or without a current. An inspection of the lattice configurations 
also reveals no clues of a phase transition at $T\simeq 0.13$. In Figure 
\ref{dielectric}, characteristic vortex configurations in the lattice
are show for various temperatures for a relatively large value of the electric
field, $E=0.3$. Even at the lowest temperature shown here ($T=0.17$), there are 
few vortices and they are clearly bound to one other. (One can also 
see in that figure that the vortices seem to become unbound above the expected
$T_c$ of 0.218.) Hence, the apparent FFH scaling collapse at the large value of 
$z$ and the small value of $T_c$ does not seem to correspond with a real phase
transition. We now investigate if the FFH scaling collapse for these values of
$T_c$ and $z$ could be due to finite size effects, as recently suggested by 
Medvedyeva {\it et al.}\cite{minnhagen00}

\begin{figure}[t]
\centerline{
\epsfig{file=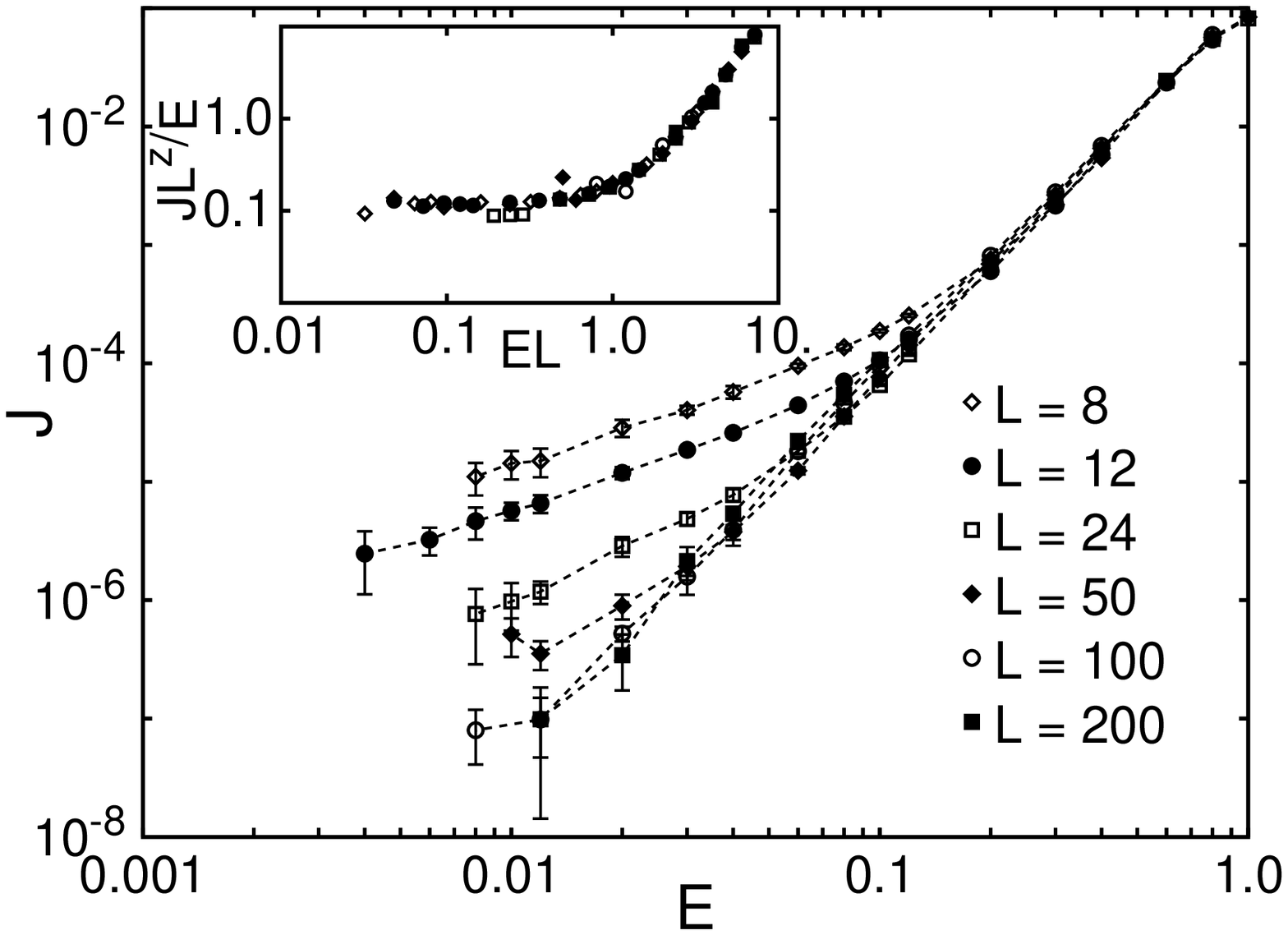,angle=0,width=3.4in}}
\caption{
The $J$-$E$ curves for $T=0.210$ for a variety of system sizes. [Inset:
The same set of data scaled with finite size theory (Eq.~\protect\ref{fsseq}) for 
$z=2.15$.]}
\label{finsiz}
\end{figure}

To further explore the value of $z$ and the transition temperature, 
we employed finite-size scaling methods for $T=0.210$, a temperature 
just below the expected value $T_c=0.218$.\cite{leeteitel92,wallin95} 
The $J$-$E$ curves for values of $L$ ranging from $12$ to $200$ are shown 
in Figure \ref{finsiz}.  As one can see, the curves become straighter 
(on a log-log scale) as expected near the critical isotherm when  
the system size is increased, indicating that the low-$E$ Ohmic behavior 
for the $L=24$ data is attributable to finite size effects and that the 
true critical isotherm occurs at a larger temperature. The finite size 
scaling, shown in the inset to Figure \ref{finsiz}, reinforces this 
idea. With $z$ being the only fitting parameter, the optimal value is 2.15, 
reinforcing the original results of Lee and Teitel\cite{leeteitel94} (and 
the subsequent work of others) that $z=2$ and that $T_c$ occurs at a much 
larger value than obtained from the FFH scaling.  The fact that the value 
of $z$ here is slightly larger than 2 can be attributed to the fact that 
the temperature is just below that of the expected value of the 
transition temperature and that $z$ increases in value for $T<T_c$.
 
\begin{figure}[t]
\centerline{
\epsfig{file=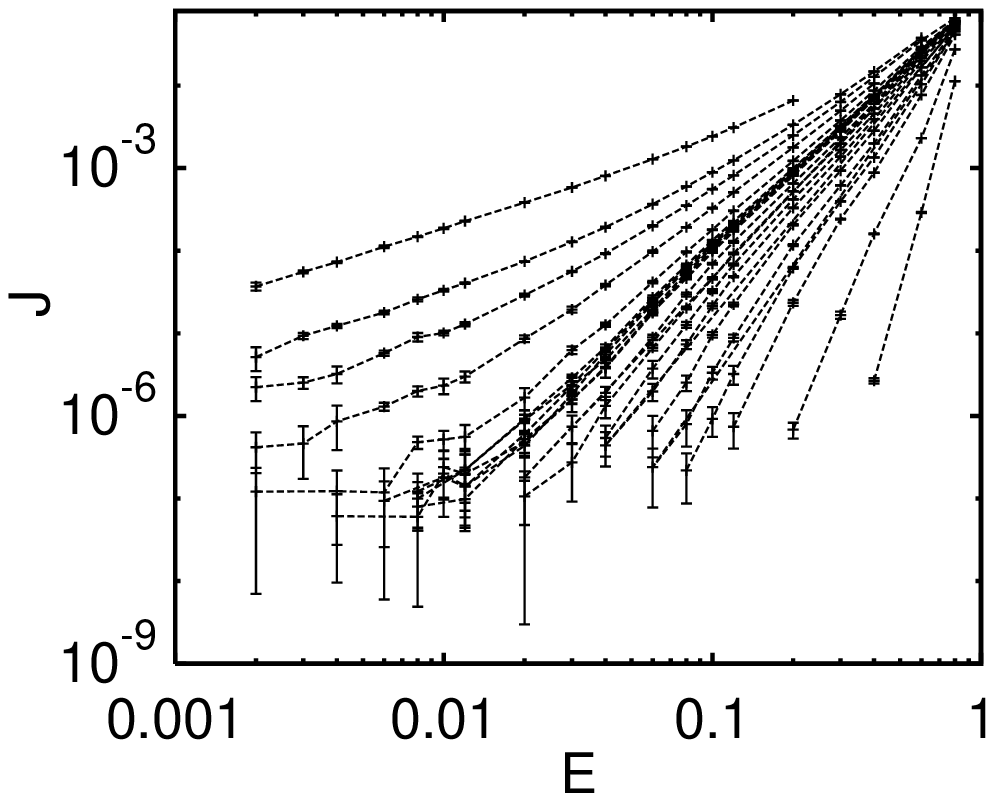,angle=0,width=3.4in}}
\caption{
Charge current density $J(E,T)$ versus $E$ for fixed system
size $L=100$ for the temperatures  0.09, 0.12, 0.15, 0.16, 0.17, 0.18, 0.19,
0.195, 0.20, 0.205, 0.21, 0.211, 0.212, 0.213, 0.214, 0.215, 0.22, 0.23, 0.24,
0.25, 0.28 (from right to left). The density of curves is larger near $T_c$.}
\label{L100raw}
\end{figure}

\begin{figure}[t]
\centerline{
\epsfig{file=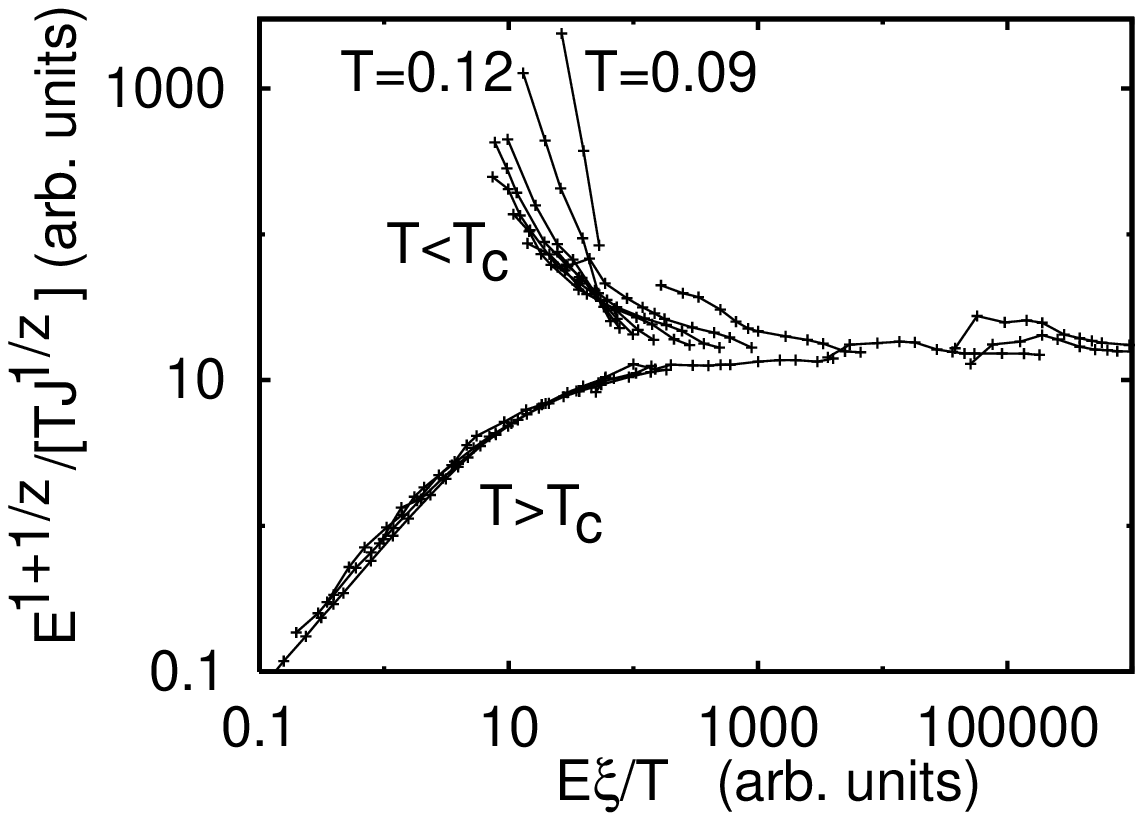,angle=0,width=3.4in}}
\caption{
The $L=100$ data of Figure \protect\ref{L100raw} scaled according
to FFH scaling (Eq.~\protect\ref{FFHsc}) using $z=1.90$, $T_c=0.212$, and $C=0.65$. 
The $T=0.09$ and $T=0.12$ isotherms do not collapse with the other curves, which we
attribute to those curves being far from the critical region.}
\label{L100FFH}
\end{figure}

The value of $z$ obtained in the finite size scaling reinforces the idea 
that the large value of $z$ found for the $L=24$ data in the FFH scaling 
is due to finite size effects. We have therefore calculated $J$-$E$ curves 
for $L=100$. (See Figure \ref{L100raw}.) The $J$-$E$ curves are similar 
to those of $L=24$ both in terms of qualitative behavior and in terms of
the error bars. The primary difference is that the intermediate-temperature 
curves are straighter than those of $L=24$ and the critical isotherm now 
occurs at a temperature $T\simeq 0.210$ for $z\simeq 2$. (Of course, 
if we could extend the $J$-$E$ data to smaller $J$, the $J$-$E$ curves 
would not be straight for $T<T_c$ due to finite size effects.) Using 
these values as a guide, the best collapse for this data with the FFH 
scaling, shown in Figure \ref{L100FFH}, was obtained using $z=1.90$, 
$T_c=0.212$, and $C=0.65$. As one can see, the collapse is quite good for 
the $T>T_c$ data. (Note that four points for $T=0.22$ were 
excluded from this figure due to the fact that the error bars were
greater than $50\%$.) Even the $T<T_c$ data scale well despite the fact that
$z$ in this regime has a temperature dependence different than
that of the scaling parameter $x$. The 
$T=0.090$ and $T=0.12$ are the exception to the collapse but is most likely due 
to the fact that that data is well outside the critical region. The fact that the rest of
the $T<T_c$ data scales reasonably well may be a coincidence or evidence for a
finite correlation length. (We present evidence for a finite correlation length for
$T<T_c$ in the next subsection.) The value of the $T_c$ achieved through our methods
is slightly smaller than the expected value, which we attribute to finite size 
effects. We add that an acceptable collapse is also achieved for $T_c=0.218$ and $z=2$. 

We believe that this work is a definitive verification of $z=2.0 \pm0.2$ found 
using FFH scaling and therefore puts FFH scaling in accord with finite size scaling.
We conclude that the larger value of $z$ (5.6) found for the $L=24$ data and 
elsewhere\cite{pierson99,ammirata99,pierson00b} is due to finite 
size effects.\cite{minnhagen00} (See discussion in Section \ref{sec:summ}.) 
This work also vindicates the use of FFH scaling for systems free of finite
size effects. Lee and Teitel\cite{leeteitel94} and Harris {\it et 
al.}\cite{harris91} have previously showed a $z=2$ collapse using FFH scaling 
for their numerical data ($L=24$) and JJA data, respectively. However, we 
do not believe that that work should not be taken as evidence for $z=2$ because 
the scaling collapse was not convincing, they did not vary $z$ to optimize the 
collapse, and it was later shown that the data collapse better for 
much larger values of $z$.\cite{pierson99,pierson00b}

Finally, we note that we calculated the values of the $J$-$E$ exponent 
$\alpha$ in order to study the predictions of Minnhagen {\it et 
al.}\cite{jonsson97} about its temperature dependence. In accord with 
the work of others,\cite{holmlund96,weber96} our work indicates that 
the prediction of Ref.~\onlinecite{jonsson97} for the temperature 
dependence of $\alpha$ (usually denoted $\alpha_{PM}$) is in better 
agreement with the data than the original prediction\cite{ahns80} 
for the temperature dependence of $\alpha$ (sometimes denoted
$\alpha_{AHNS}$.) This agreement further verifies our 
$L=100$ $J$-$E$ data.

\subsection{Vortex Correlation Length for $T<T_c$}

Our method for studying the vortex correlation length is to investigate the 
value of the $J$-$E$ exponent $\alpha$ as a function of $E$ by looking at 
the slope of our $J$-$E$ curves. More specifically, we study the derivative 
$d\log J/d\log E$, which is equal to $\alpha$ when $J\propto E^\alpha$. 
A similar examination by Medvedyeva {\it et al.}\cite{minnhagen00} (see 
their Figure 2) for system sizes of $L=8$, using the analogous $V$-$I$ 
curves in the 2D RSJ model, revealed that the slope started at a low 
value for small $I$, peaked, and then descended to a value near 1 at 
large values of $I$. For each of the isotherms with temperatures below 
the critical temperature, they found the peak to occur at roughly the 
same value of $I$.\cite{minnhagen00} Because the scaling variable is 
either $I\xi$  when $\xi\ll L$ or $IL$ when $L\lesssim\xi$, the fact 
that the peak in $d\log V/d\log I$ does not move is an indication that 
finite size events dominate this temperature regime (and the scaling 
parameter is $EL$). Above $T_c$, the authors\cite{minnhagen00} found the 
peak to move quickly to larger values of $I$, indicating that $I\xi$ is 
the principal scaling variable since $\xi$ is decreasing quickly in this regime.

\begin{figure}[t]
\centerline{
\epsfig{file=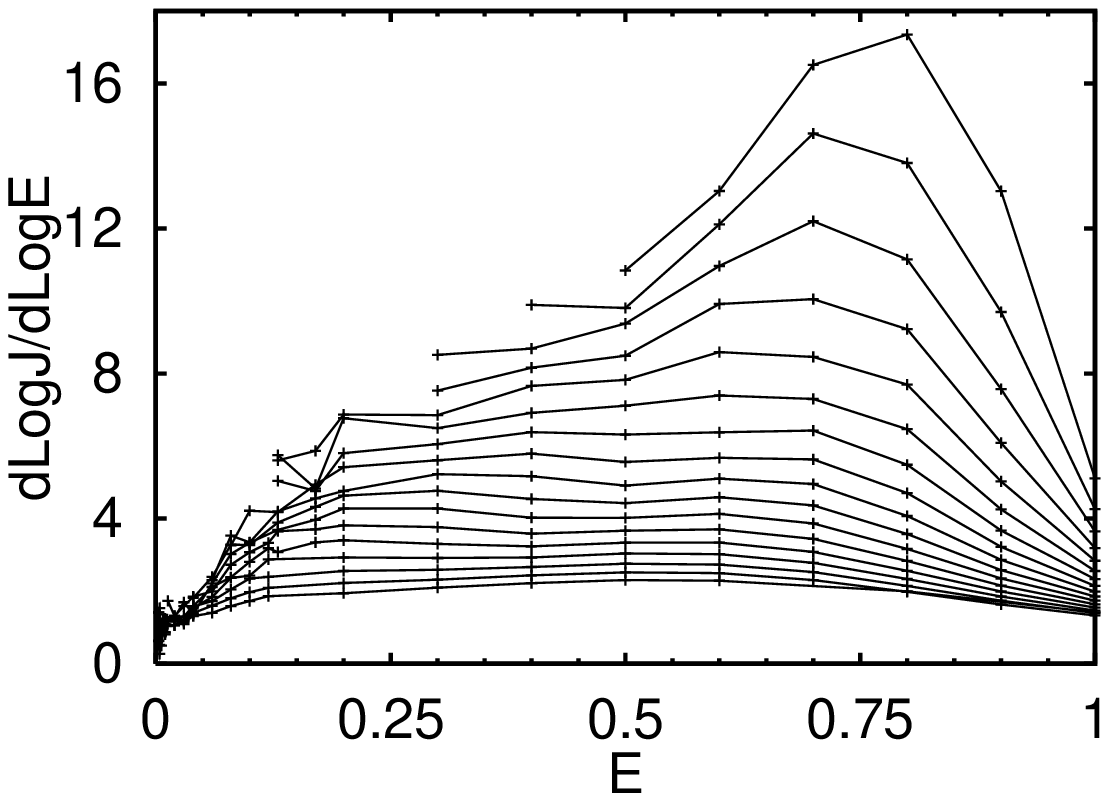,angle=0,width=3.4in}}
\caption{
The $L=24$ $J$-$E$ data plotted as $d\log J/d\log E$ versus $E$ for
all of the temperatures shown in Figure \protect\ref{L24raw}. The top curve corresponds
to $T=0.09$ and the bottom curve to $T=0.25$.}
\label{slope} 
\end{figure}

For our $L=24$ data, $d\log J/d\log E$ behaves differently below $T_c$ 
than does the analogous quantity of Ref.~\onlinecite{minnhagen00}. As 
one can see in Figure \ref{slope}, the locations of the peaks as a function 
of $E$ move to smaller values as one increases the temperature from 0.09 
to about 0.14. In the temperature range 0.15-0.22, the peaks develop into 
plateaus. 

The peaks and plateaus indicate that there are two competing length scales. 
On the large length scale side (small $E$) is the temperature-dependent 
correlation length $\xi_-(T)$ while at small length scales (large $E$)
is the lattice size, which enters because of a saturation of vortex density. 
A plateau signifies that $\xi$ is much larger than the saturation length 
scale while a peak indicates that 
$\xi$ is approaching the lattice size. The value for the left hand size of 
the plateau is determined by $\xi$ or $L$, whichever is smaller. In Figure 
\ref{slope}, the point at which $\xi$ becomes larger than $L$ occurs at 
about $T=0.19$, above which the plateau begins at $E\simeq 0.13$ on the 
low $E$ side. That the value of $E$ for which $d\log J/d\log E$ peaks or begins to 
plateau depends on temperature for a broad temperature range below 
$T_c$ is an indication that the scaling variable for $T<0.19<T_c$ is 
$E\xi$ and that, therefore, $\xi_-(T)$ is finite below $T_c$. 

The lattice spacing affects the $J$-$E$ curves for $E\gtrsim 0.7$, as can 
be seen in Figure \ref{slope}, and determines a maximum value of $E$ at 
which $d\log J/d\log E$ peaks. Therefore, the discrete lattice size can
also result in the position of the peak in $d\log J/d\log E$ not moving. 
As a consequence, that the peaks would be located all at the same approximate 
value of $E$, as found in Ref.~\onlinecite{minnhagen00}, should not 
be taken as evidence strictly for finite size effects. 

The values of $E$ that correspond to the influence of the lattice spacing
and the lattice size can be predicted by simple arguments. (See for example
Ref.~\onlinecite{pierson99}.) The approximate energy for a vortex pair is 
$(\log R -ER)/\epsilon$, which increases for small $R$ and then 
decreases for large $R$, thereby peaking at $R=1/E$.\cite{leeteitel94} 
(This equation explains 
why one probes different length scales by changing the value of $E$ as 
discussed in Section IIIA.) Since the lattice size corresponds to a value $1$,
one expects its effects to manifest themselves for $E\simeq 1$, which is what
we observe in Figure \ref{slope}. Similarly, one expects finite size effects to 
manifest themselves for $E\simeq 1/L\simeq 0.04$. Indeed, in Figure \ref{slope}, 
all of the isotherms are ohmic for $E\lesssim 0.5$, in agreement with our prediction.

\section{Discussion and Summary}    
\label{sec:summ} 

Using system sizes up to $L=200$ in Monte Carlo simulations 
of the 2DCG, we conclude $z=2.0\pm0.2$ using Fisher-Fisher-Huse scaling 
and finite size scaling, reinforcing the results of others that $z=2$. 
(See, e.g., Refs.~\onlinecite{minnhagen00,leeteitel94,jose97}.) We have 
also presented evidence that $\xi_-$, the correlation length below 
$T_c$, is finite.  

Previous analysis of experimental data on 2D superconductors, 
JJA's, and superfluids, and of numerical data on the 2DCG by 
two of these authors\cite{pierson99,ammirata99,pierson00b} 
using FFH scaling indicated $z\simeq 5.6$. As stated here and 
in Ref.~\onlinecite{minnhagen00}, it appears that this large 
value is due to finite size effects. Because a majority of 
the $V$-$I$ data from the literature was examined in 
Ref.~\onlinecite{pierson99}, it appears that nearly all 
experimental $V$-$I$ measurements on 2D superconductors 
and JJA's are dominated by finite size effects. Why the value 
$z\sim 6$ should result from finite size effects\cite{minnhagen00} 
remains a mystery, and should be studied further.

Because these results indicate that finite size effects are 
ubiquitous in 2D superconductors and 2D JJA's, the $V$-$I$ 
and $R$ data from the literature that were used to 
verify KTB behavior in the conventional manner needs to be 
re-examined to account for the finite size effects.  In the 
presence of strong finite size effects like 
those in these samples, two aspects of the conventional method fail. 
First, the small $I$ behavior of the $V$-$I$ isotherms would not 
reveal any quick changes in $\alpha$ near $T_c$ because finite size 
effects dominate at small $I$. In other words, as is well known, 
finite size effects wash out the expected jump in $\alpha$. Secondly 
and more importantly, the resistance formulas are no longer valid in 
the presence of finite size effects. That previous authors could fit 
their resistance data to the more commonly used Minnhagen formula is 
most likely due to the number of fitting parameters. Finally, we 
stress that the most effective means for studying the temperature
dependence of $\alpha$ should begin by extracting $\alpha(T)$ from the 
$V$-$I$ curves in 2D superconductors and JJA's for the same value of $I$.

Clearly, in order for this data to be re-interpreted, a better 
understanding of finite size effects is needed since previous checks
for finite size effects in the data of Ref.~\onlinecite{repaci96} were
negative. For example, those authors\cite{repaci96} found that the 
value of $\lambda_\perp(0)$ derived from a fit to $\lambda_\perp(T)$ 
was four times larger than the expected value. Furthermore, for that 
same data,\cite{repaci96} where finite size effects are now 
known to dominate, the resistance was found to fit better to the 
Kosterlitz resistance formula than a resistance formula derived for 
finite-size dominated Ohmic resistance.\cite{pierson99} Finally, in 
Ref.~\onlinecite{pierson99}, it was found that the length scale at 
which the crossover from finite-size-induced Ohmic behavior to 
non-Ohmic behavior was expected to occur for the data of 
Ref.~\onlinecite{repaci96} did not correspond with the observed value. 

\acknowledgements    

Acknowledgement is made by SWP to the donors of The Petroleum Research 
Fund, administered by the ACS, for support of this research.

\end{document}